\documentclass{article}

\makeatletter
\renewcommand{\maketitle}{%
  \begingroup
    \renewcommand{\thefootnote}{\fnsymbol{footnote}}%
    \if@twocolumn
      \@topnewpage[\@maketitle]
    \else
      \newpage
      \global\@topnum\z@
      \@maketitle
    \fi
    \thispagestyle{empty}
    \@thanks
  \endgroup
  \setcounter{footnote}{0}%
}

\renewcommand{\@maketitle}{%
  \newpage
  \null
  \vskip 2em%
  \begin{center}%
    {\LARGE \@title \par}%
    \vskip 1em%
    {\large \lineskip .5em%
      \begin{tabular}[t]{c}%
        \@author
      \end{tabular}\par}%
    \vskip 1em%
    {\large \@date}%
  \end{center}%
  \vskip 2em%
}
\makeatother

\usepackage[utf8]{inputenc}
\usepackage[T1]{fontenc} 
\usepackage{xcolor}
\definecolor{darkblue}{RGB}{0, 0, 100}

\usepackage{hyperref}
\hypersetup{
    colorlinks=true, 
    linkcolor=darkblue,
    citecolor=darkblue,
    urlcolor=darkblue
}

\usepackage{arxiv}
\usepackage{url} 
\usepackage{booktabs} 
\usepackage{amsfonts} 
\usepackage{nicefrac} 
\usepackage{amsmath} 
\usepackage{microtype} 
\usepackage{cleveref}  
\usepackage{graphicx}
\usepackage{natbib}
\usepackage{doi}
\usepackage{threeparttable}

\usepackage{amssymb}
\usepackage{amsthm}

\usepackage{subcaption}

\usepackage{setspace}
\usepackage{tabularx}

\theoremstyle{plain}
\newtheorem{proposition}{Proposition}

\theoremstyle{definition}

\newcommand{\Ke}{K_{\mathrm e}}
\newcommand{\gK}{g(K)}
\newcommand{\tAnaive}{t_{A}^{*}}       
\newcommand{\tAinf}{t_{A}^{\dagger}}   
\newcommand{\tAinfC}{t_{A}^{\ddagger}} 
\newcommand{\Mismatch}{\operatorname{Mismatch}}

\usepackage{fancyhdr}
\usepackage[explicit]{titlesec}
\titleformat{\section}{\large\bfseries}{\thesection}{1em}{#1}

\usepackage{titling}
\usepackage{mathpazo}

\pretitle{\begin{center}\LARGE\bfseries}
\posttitle{\par\end{center}\vskip 0.5em}
\preauthor{\begin{center}\large}
\postauthor{\end{center}}
\predate{\begin{center}\small}
\postdate{\end{center}}

\usepackage{lineno}
\begin{document}

\title{Training for Obsolescence? \\ The AI-Driven Education Trap}

\author{\href{https://orcid.org/0000-0002-0811-3515}{\includegraphics[scale=0.06]{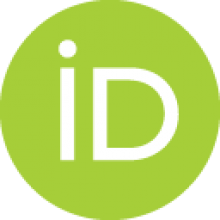}} \hspace{1mm} Andrew J. Peterson
 \thanks{Assistant Professor (Ma\^{i}tre de conf\'{e}rences), University of Poitiers. \href{mailto:andrew.peterson@univ-poitiers.fr}{andrew.peterson@univ-poitiers.fr}}}

\renewcommand{\shorttitle}{Training for Obsolescence? }

\hypersetup{
pdftitle={Training for Obsolescence? The AI-Driven Education Trap}
pdfsubject={AI disruption of the workplace, skill mismatch, AI impact on education},
pdfauthor={Andrew J.~Peterson},
pdfkeywords={Artificial Intelligence,  Skill Mismatch, Human Capital , Coordination Failure, Non-cognitive skills , General-Purpose Technology},
}

\date{\today}

\maketitle

\bigskip

\begin{abstract}

Artificial intelligence is simultaneously transforming the production function of human capital in schools and the return to skills in the labor market. We develop a theoretical model to analyze the potential for misallocation when these two forces are considered in isolation. We study an educational planner who observes AI's immediate productivity benefits in teaching specific skills but fails to fully internalize the technology's future wage-suppressing effects on those same skills. Motivated by a pre-registered pilot study suggesting a positive correlation between a skill's "teachability" by AI and its vulnerability to automation, we show that this information friction leads to a systematic skill mismatch. The planner over-invests in skills destined for obsolescence, a distortion that increases monotonically with AI prevalence. Extensions demonstrate that this mismatch is exacerbated by the neglect of unpriced non-cognitive skills and by the endogenous over-adoption of educational technology. 
Our findings caution that policies promoting AI in education, if not paired with forward-looking labor market signals, may paradoxically undermine students' long-term human capital, such as by crowding out skills like persistence that are forged through intellectual struggle.

\end{abstract}

\bigskip

JEL Codes: J24, I28, O33, J31, D91

\section{Introduction}

The rapid advancement of artificial intelligence presents a unique challenge to the economics of human capital. As a general-purpose technology, AI is poised to simultaneously transform the production of human capital in schools and alter the demand for it in the labor market \citep{Agrawal2018, acemoglu_2018race}. While the literature has extensively analyzed these effects in isolation, focusing either on EdTech's potential to boost learning outcomes \citep{escalada2023} or on automation's threat to wages \citep{acemoglu_restoration_2019}, their interaction remains underexplored. This separation obscures a critical policy dilemma: the very features that make a skill easier to teach with AI (e.g., codifiability, clear rules) may be the same features that make it susceptible to automation.
This creates a technology-driven wedge between the production of skills and their economic return, potentially incentivizing educational systems to specialize in precisely the wrong areas.

This paper develops a theoretical framework to analyze how this dual impact of AI can generate an economically inefficient skill mismatch. 
We model the decision problem of an educational planner who faces a trade-off between investing in a skill that is increasingly easy to produce (due to AI-assisted teaching) but whose market value may erode (due to AI-driven substitution). 
The core friction is behavioral: the planner is guided by the salient, immediate heuristic of classroom productivity and fails to fully internalize the complex, future general equilibrium effects on wages.
A wedge arises because the planner observes the positive shock to the education production function but neglects the negative shock to the labor demand function.
Our central research question is therefore: \textit{How does AI's asymmetric impact on the technology of skill formation and on wage schedules leads myopic educational planners to tilt the human-capital portfolio toward AI-susceptible skills and away from AI-robust and harder-to-measure dimensions of human capital?}

To formalize this, we develop a two-stage model of skill acquisition and labor market competition. 
The model's central mechanism is motivated by the observation, supported by a pre-registered pilot survey (detailed in the Online Appendix), that skills made easier to teach by AI are often the same ones it devalues in the workplace. 
The framework isolates two opposing channels: an \emph{education channel}, where AI acts as a complement to learning time, and a \emph{substitution channel}, where AI acts as a substitute for human labor.
We demonstrate that a ``naive'' planner, reacting only to the education channel, systematically over-invests in skills destined for obsolescence compared to an ``informed'' planner who anticipates the substitution channel.
This behavior creates a mismatch that is not merely a static error but a dynamic trap: under standard conditions, the misallocation of human capital monotonically increases with the level of AI technology.

Our baseline model allows for several extensions that demonstrate the robustness of this core mechanism. 
First, we show that this misallocation is amplified when we account for unpriced, non-cognitive skills. Because these skills (e.g., persistence) are often jointly produced through the labor-intensive learning processes that AI aims to reduce, a naive planner's optimization inadvertently crowds out their development. 
Second, we endogenize the costly adoption of educational AI and find a bias toward over-investing in technologies that enhance easily measured skills at the expense of these unpriced dimensions.
Finally, motivated by debates over skills like coding, where AI may simultaneously increase the utility of basic literacy while devaluing intermediate proficiency, we explore non-monotonic wage effects. We show how a naive planner can fall into ``substitution traps,'' directing students into a middle-skill bracket that is maximally exposed to automation.

This paper contributes to the literature on human capital formation by explicitly linking the technology of skill production with the technology of skill demand. We build on the task-based framework of labor economics \citep{acemoglu_autor2011} and connect it to the literature on education production functions \citep{hanushek2020_edprod}. Our work also relates to models of directed technical change \citep{acemoglu2002directed}, applying the logic of induced innovation to the educational sector's technology adoption decisions. 
Our contribution is to formalize a novel mechanism generating human capital mismatch: not a failure to anticipate aggregate supply responses (as in classic cobweb models), but a failure to integrate the cross-domain impacts of a single technological shock.

The remainder of the paper is structured as follows. Section~\ref{sec:motivation} reviews the motivation for our approach and the relevant literature. 
Section~\ref{sec:model} develops our baseline model to isolate the core education trap, then presents several extensions that analyze the role of non-cognitive skills, endogenous technology adoption, and non-monotonic wage effects. 
Section~\ref{sec:discussion_conclusion} discusses policy implications and concludes.


\section{Motivation and Related Literature}\label{sec:motivation}

\subsection{AI as a General-Purpose Technology and the Mismatch Problem}

Our central contribution is to formally connect two bodies of literature that have largely developed in parallel: the impact of artificial intelligence (AI) on labor market demand and its impact on the educational production function. We structure this review to directly motivate our model's core assumptions and extensions. First, we review the literature on AI's substitutive effects on labor to ground our assumptions about wage pressure. Second, we survey the evidence on AI's productivity-enhancing role in education. Finally, we connect our approach to classic models of educational choice under information frictions and motivate the model's key extensions concerning unpriced skills and endogenous technology adoption.

\subsection{The Labor Market Channel: Skill Substitution and Wage Effects}

Research on technological change has progressed from a canonical skill-biased view to a task-based framework that treats jobs as bundles of activities with varying susceptibility to automation \citep{acemoglu_autor2011}. Within this framework, AI acts as a general-purpose technology whose frontier algorithms substitute for routine, codifiable cognition while complementing non-routine analytical, creative, and socio-emotional tasks \citep{autor2015automation}. Formal models decompose AI’s impact into a \emph{displacement effect}, which substitutes for human labor in existing tasks, and a \emph{productivity effect}, which lowers costs, expands output, and creates new human-centered tasks \citep{acemoglu_automation_2019, bessen2019automation}.

Empirical evidence confirms this heterogeneity. Patent-based exposure indices show that tasks vulnerable to current AI are concentrated in clerical, administrative, and certain analytical occupations, whereas tasks intensive in problem-solving or interaction remain comparatively insulated \citep{webb2019impact, felten_raj_seamans_2021}. 
Studies of industrial robots estimate significant wage losses borne by routine task specialists \citep{acemoglu_robots_2020}. 
Collectively, these findings indicate that AI capital exerts downward wage pressure on automatable skills while potentially raising the marginal product of complementary skills. This literature often treats the supply of skills as given or slow-moving, leaving the mechanisms of educational response under-theorized, which directly motivates our assumption of \emph{asymmetric workplace substitution}.

\subsection{The Education Channel: Asymmetric Productivity Gains}

Within the framework of education production functions \citep{hanushek2020_edprod}, a rapidly growing literature documents AI's capacity to lift teaching productivity, particularly where learning objectives are structured, rule-based, and easily assessed \citep{fazlollahi2022effect}. While the effects of education technology are often mixed \citep{escueta2017education}, compelling evidence for gains in structured domains predates the current AI wave. A meta-analysis found intelligent tutoring systems were already achieving effectiveness nearly on par with one-on-one human tutoring \citep{vanlehn2011relative}, a benchmark of high pedagogical effectiveness \citep{bloom_2sigma1984}. Recent advances in generative AI appear to amplify these efficiencies. Randomized trials demonstrate dramatic learning gains from AI tutors in subjects like mathematics and English, sometimes equivalent to one to two years of conventional schooling, in both high-resource and developing-country settings \citep{kestin2024_tutoring, Simone_2025chalkboards, henkel_2024_math}.

However, a crucial strand of this research provides a cautionary note, showing that such pedagogical efficiency does not guarantee deeper or more durable learning \citep{gerlich2025ai, carter2017impact}, and that EdTech has been shown to focus disproportionately on STEM \cite{alam_foundation_2022}. 
These heterogeneous returns provide a strong incentive for educators, especially those evaluated on narrow, quantitative metrics, to expand AI-mediated instruction where its effects are most visible. This empirical regularity supplies the microfoundation for our assumption of \emph{asymmetric AI complementarity in education}, which posits that AI enhances the marginal product of teaching time for certain kinds of skills more than others.

\subsection{Modeling Educational Choice under Uncertainty and Information Frictions}
The core friction in our model is rooted in the challenge of making human-capital decisions amidst rapid technological change, a modern incarnation of the classic ``race between education and technology'' \citep{goldin2008race, autor2020extending}. Skill mismatch, that is the divergence between competencies supplied by education and those demanded by firms, carries sizable welfare losses \citep{quintini2011mismatch, brunello2021skill}. 
Classic ``cobweb'' models show how enrollment decisions that naïvely extrapolate current wage premia produce cyclical skill imbalances \citep{freeman1976cobweb, ryoo2004engineering}. This problem is one of decision-making under severe uncertainty, where point forecasts of future returns are unreliable guides for policy \citep{manski2004measuring, frank2019toward}.
Our model builds on this cobweb tradition but identifies a critically different source of friction. 
In the classic framework, the forecasting error stems from a failure to anticipate the \emph{endogenous supply response} of other agents to a public price signal. 
By contrast, the mechanism we propose is driven by a failure to anticipate the dual, cross-domain consequences of a single technological shock. 
The planner's error is not a failure to predict the actions of their peers, but a failure to connect AI's immediate, positive productivity effect in education with its future, negative substitution effect in the labor market. 
This behavioral assumption is consistent with institutional incentives: educational planners are often evaluated on immediate, measurable learning outcomes (e.g., test scores), making them rationally responsive to local productivity signals while underweighting distant and unpriced labor-market externalities.
This distinction motivates our central modeling choice: contrasting a planner who reacts only to the immediate productivity signal with a fully-informed planner who anticipates AI's complete general equilibrium impact on wages.  
The divergence in their choices provides a novel channel for how a general-purpose technology can systematically generate human capital misallocation, a friction distinct from those in the canonical literature.

\subsection{Displacement of Non-Cognitive Skills as a Negative Externality}

Our first extension introduces a critical, often unpriced, dimension of human capital, reflecting a foundational tension in educational philosophy: the risk of prioritizing narrow vocational training over the holistic development of citizens \citep{dewey2024democracy}. More recent literature confirms that the social value of education extends far beyond wages. Schooling generates substantial non-pecuniary benefits, including better health, more stable families, and higher levels of social trust, that represent a significant, if unpriced, return on educational investment \citep{oreopoulos2011priceless}.

The economic literature has documented the immense and growing labor market value of these non-cognitive skills \citep{deming2017growing, lundberg2017non}, which are often poorly measured and inadequately incentivized within education systems that prioritize cognitive assessments \citep{heckman2012hard, heckman2013fostering}. The intense focus on AI-teachable skills (our skill $A$) thus risks displacing the development of these competencies, creating a negative externality by degrading the production of a valuable social good with proven long-run benefits \citep{jackson2018test, peterson2025ai}.

Recent research provides a specific mechanism for this, showing that reliance on AI can promote ``cognitive offloading,'' which undermines the deep processing required for durable learning \citep{jose2025cognitive}. For instance, evidence suggests that using AI as a ``crutch" for problem-solving can inhibit the development of underlying conceptual understanding, with students performing worse when the tool is removed \citep{kosmyna2025your}. 
This aligns with findings that frequent AI tool usage can be negatively correlated with critical thinking abilities, mediated by this increase in cognitive offloading \citep{gerlich2025ai}.  
Consequently, the net effect of AI may depend on whether it is used to develop metacognitive skills like self-regulation or merely to find efficient shortcuts \citep{zhou2024mediating}, that might even harm non-cognitive skills. 
This dynamic provides a clear illustration of a classic multitasking problem \citep{holmstrom1991multitask}: when a planner's focus is drawn to an easily measured task (teaching skill $A$), the introduction of a technology that boosts its productivity will predictably divert resources from other valuable but unmeasured tasks. The extension in Section~\ref{sec:noncogn-extension} of our model formalizes how this distortion amplifies the social cost of the initial mismatch.

\subsection{Endogenous Technology Adoption and Directed Change}
This threat to unpriced virtues becomes particularly acute when we consider the active, costly choice of AI adoption by educational institutions. We frame this within the logic of principal-agent problems, where the decision-maker (e.g., a school administrator) optimizes over a narrow set of objectives. A school leader, assessing which tasks have high \emph{suitability for machine learning} \citep{brynjolfsson2018what} may choose to adopt an AI teaching tool based on its documented ability to improve test scores in skill $A$. 
However, this ignores negative externalities on the development of other, unpriced skills (for contextual evidence on unintended consequences of rapid ed-tech adoption, see \citealt{west2023ed}). 
This logic mirrors the framework of directed technical change\citep{acemoglu2002directed}, where innovation is endogenously directed toward factors with higher perceived (and privately captured) profitability, even if this leads to socially suboptimal outcomes. This underpins an extension  of our model (Section~\ref{sec:ai_intensity})
 where the naive planner's chosen AI intensity ($\alpha^*$) is inefficiently high because it ignores this collateral social cost, a specific case of the general finding that decentralized adoption of transformative technologies with unpriced risks tends to be too fast \citep{acemoglu2024regulating}.

\subsection{Non-Monotonic Returns and Within-Skill Polarization}
Finally, our extension of non-monotonic skill returns refines the canonical task-based model of labor market polarization\citep{acemoglu_autor2011}. While existing models effectively explain the `hollowing out' of the occupational structure by skill level, our extension examines how automation can create a ``barbell effect'' \emph{within} a skill category. Recent theory provides a direct foundation for this mechanism; for instance, \cite{acemoglu2022automation} show that when automation targets tasks of ``interior" or medium complexity, it naturally generates polarization by pushing human labor to the simplest and most advanced tasks. This aligns with firm-level evidence demonstrating that generative AI disproportionately boosts the productivity of lower-skilled agents while having little effect on top performers, effectively creating a non-linear return to skill within the same role \citep{brynjolfsson_generative_2023}. Our model formalizes this dynamic, showing how a naive planner might inefficiently over-produce for the middle-skill tier. 

\subsection{The Model's core assumption}

The preceding literature review highlights a critical tension: the same attributes that make a skill suitable for AI-assisted teaching (e.g., being rule-based, codifiable) often make it vulnerable to automation in the workplace. Our model's central mechanism rests on formalizing this tension as a core assumption of a positive correlation between AI's effectiveness as a teaching tool and its substitutive pressure on wages for a given skill.

This assumption is grounded in the task-based framework and is consistent with preliminary evidence from a pre-registered pilot study we conducted.\footnote{We conducted a pre-registered, exploratory survey (n=20) to gauge initial perceptions of this relationship. The results, which show a positive correlation and are presented in an Online Appendix, are intended for illustrative purposes only. The model's formal validity does not depend on these preliminary findings, which are also available on our \href{https://osf.io/nwy4c/?view_only=25b2a80ac79a44b890284dde2d70b8b8}{OSF repository}.}  While not a formal test of our model, the survey's qualitative findings provided the initial impetus for exploring the coordination failure that we formalize in the subsequent sections.

\bigskip

\section{Model}\label{sec:model}

We develop a model to analyze the tension between two countervailing effects of AI on human capital formation. The first is an education channel, where AI enhances the productivity of acquiring certain skills. The second is a workplace substitution channel, where AI substitutes for those same skills in the labor market. 

Consider a representative student with a time endowment $T$ to be allocated between learning two skills, $A$ and $B$. Skill acquisition depends on time investment $t_j$ and the level of educational AI, $\Ke$, which tracks the general level of AI in the economy, $K$. The production functions $A = f_A(t_A, \Ke)$ and $B = f_B(t_B, \Ke)$ are assumed to be strictly increasing and strictly concave in time.

We compare the time allocation decisions of two planners. A \emph{naive planner} (e.g., a school administrator or student) observes that AI makes learning skill $A$ more efficient and maximizes skill output valued at current, fixed prices. An \emph{informed planner} anticipates that AI will also depreciate the value of skill $A$ in the labor market and maximizes the future economic value of skills.

\subsection{The Naive Planner: The Education Channel}

The naive planner observes fixed price signals $p_A$ and $p_B$ (e.g., current wages) and chooses $t_A$ to maximize:
\[
    U_{\text{naive}}(t_A; K) = p_A f_A(t_A, g(K)) + p_B f_B(T-t_A, g(K)).
\]
We assume that AI is relatively more complementary to teaching skill $A$. Formally, the marginal return to time spent on skill $A$ increases with AI relative to skill $B$, implying that $U_{\text{naive}}$ has increasing differences in $(t_A, K)$.\footnote{If $f_j$ are differentiable, this is equivalent to $\frac{\partial^2 U_{\text{naive}}}{\partial t_A \partial K} > 0$.}
It follows immediately from Topkis's Theorem that the naive planner's optimal allocation, $\tAnaive(K)$, is strictly increasing in the level of AI. As AI tools improve, the naive planner doubles down on the skill that is becoming easier to teach.

\subsection{The Informed Planner: The Labor Market Channel}

The informed planner anticipates equilibrium wages $w_A(K)$ and $w_B(K)$. We assume AI is a labor substitute for skill $A$, so $w_A(K)$ declines relative to $w_B(K)$ as $K$ increases. The informed planner maximizes:
\[
    U_{\text{inf}}(t_A; K) = w_A(K) f_A(t_A, g(K)) + w_B(K) f_B(T-t_A, g(K)).
\]
The key tension is that while AI facilitates \emph{learning} skill $A$, it reduces its \emph{earning} potential. We assume the labor market substitution effect dominates the pedagogical efficiency gain. This assumption rests on the divergence of marginal costs: while AI in education reduces the time required to acquire a skill, the investment remains tethered to the high opportunity cost of human attention. Conversely, AI in the labor market drives the wage for that skill toward the marginal cost of compute. Since the cost of compute falls significantly faster than the opportunity cost of human time, the asset value of the skill depreciates faster than the cost of acquiring it falls.

Consequently, the objective $U_{\text{inf}}$ exhibits \emph{decreasing differences} in $(t_A, K)$.
In contrast to the naive planner, the informed planner's optimal allocation $\tAinf(K)$ is decreasing in $K$.

\subsection{Skill Mismatch}

We define \emph{skill mismatch} as the divergence between the naive and informed allocations: $\Mismatch(K) = \tAnaive(K) - \tAinf(K)$.
Assuming the planners start from an identical allocation at some baseline $K_0$ (where fixed prices match initial wages), their paths immediately diverge.

\begin{proposition}[Growing Mismatch]\label{prop:mismatchK}
For all $K > K_0$, the naive planner invests more time in skill $A$ than is socially optimal ($\tAnaive > \tAinf$). Moreover, this mismatch is strictly increasing in the level of AI.
\end{proposition}

The intuition is straightforward: the naive planner chases the efficiency gains in teaching $A$ while ignoring the eroding market value of that skill ( proof in Appendix Section~\ref{proof:prop1mismatchK}).

\subsection{Extension: Unpriced Non-Cognitive Skills}
\label{sec:noncogn-extension}

Schools also produce unpriced non-cognitive skills, $C$, such as perseverance or social skills. We assume these skills are better formed through the labor-intensive learning of skill $B$ rather than the AI-assisted learning of skill $A$. Furthermore, reliance on AI tools may directly crowd out non-cognitive development. 
Let the social value of these skills be $\Gamma C(t_A, K)$. The informed planner now maximizes total welfare $U_{\text{inf}} + \Gamma C$. 
Since shifting time to $A$ and increasing AI reliance both harm $C$, the social planner has an even stronger incentive to reduce $t_A$. The naive planner, ignoring $C$, continues to increase $t_A$.

\begin{proposition}[Non-Cognitive Deficit]\label{thm:NoncogGap}
The presence of unpriced non-cognitive skills amplifies the mismatch. The naive planner not only over-supplies the obsolete skill $A$ but also increasingly under-provides non-cognitive skills relative to the social optimum.
\end{proposition}

The proof is provided in Appendix Section~\ref{app:proof_noncog}.

\subsection{Extension: Endogenous AI Adoption}
\label{sec:ai_intensity}

Finally, consider the decision to adopt AI tools (intensity $\alpha$) at a cost $c$. The naive planner perceives a higher marginal benefit from adoption than the social planner for two reasons: they overestimate the value of skill $A$, and they ignore the negative externality on non-cognitive skills.

\begin{proposition}[Over-Adoption]\label{prop:ai_cost}
The naive planner adopts AI tools at a strictly higher intensity than the social planner. As the cost of AI falls, this adoption gap widens, further exacerbating the skill mismatch.
\end{proposition}

See Appendix Section~\ref{proof:prop3_final} for the proof.

\subsection{Extension: Non-Monotonic Returns and the ``Substitution Trap''}
\label{sec:extension-nonmonotonic}

In our baseline model, we assumed AI is a straightforward substitute for skill $A$. We now relax this assumption to explore a scenario where the returns to skill $A$ are non-monotonic. 
We conceptualize a three-tiered wage structure:\\
1. \emph{Basic Literacy:} A low level of skill $A$ remains valuable for overseeing AI tools.\\
2. \emph{Substitution Trap:} Intermediate skill levels are directly automated by AI, yielding low returns.\\
3. \emph{Advanced Expertise:} High-level expertise remains complementary to AI, commanding a significant premium.

Formally, we model this as a piecewise wage schedule $w_A(A; K)$ with a valley in the middle. 
While the naive planner continues to optimize against a linear price signal (increasing time $t_A$ as AI makes teaching easier), the informed planner faces a discrete choice: either target basic literacy (low $t_A$) or aim for advanced expertise (high $t_A$).

\subsubsection{The Barbell Strategy}

Because the intermediate skill level yields the lowest returns, the informed planner's optimal strategy is a ``barbell'' approach: avoid the middle ground entirely.
As AI capital $K$ increases, the time required to reach the advanced threshold decreases (due to better teaching tools), but the wage premium for expertise may also fluctuate.
The informed planner will only target advanced expertise if the wage premium is sufficiently high to justify the opportunity cost of foregone skill $B$. Otherwise, they revert to basic literacy.

\begin{proposition}[The Substitution Trap]\label{prop:nonmonotonic}
In a non-monotonic wage environment, the naive planner's linear heuristic leads to a particularly damaging form of mismatch. By incrementally increasing time investment $t_A$, the naive planner steers students directly into the ``substitution trap'' by directing some to the intermediate skill range that is maximally exposed to automation. The informed planner, by contrast, would strategically avoid this region, choosing either basic literacy or advanced expertise.
\end{proposition}

This result cautions against one-size-fits-all educational mandates. A linear increase in STEM education, for instance, might strand students in an unproductive middle ground of coding ability that is easily automated, whereas a socially optimal policy would encourage either broad digital literacy or deep, specialized expertise.


\bigskip

\section{Discussion and Conclusion}
\label{sec:discussion_conclusion}

This paper formalizes an `AI-Driven Education Trap,' an institutional coordination failure where the technology that boosts teaching productivity for a given skill simultaneously erodes that skill's market value. Our central result (Proposition~\ref{prop:mismatchK}) is that the resulting skill mismatch grows with AI prevalence when educational planners operate with an information wedge (relying on misaligned prices) and an incentive wedge (ignoring unpriced externalities). This failure is institutional, not behavioral: the planner's allocation is rational under accountability regimes that reward near-term, measured outputs, even as it undermines students' long-term market value. Our model is motivated by the empirical finding that skills amenable to AI-assisted teaching often correlate with those vulnerable to workplace substitution.

The net societal impact of AI in education can be decomposed into three competing channels, clarifying the precise points of policy leverage. A change in social surplus, $\Delta W$, from an increase in AI capital can be conceptualized as:
$$
\Delta W(K)\;=\; \underbrace{\Delta \Pi_E(K)}_{\text{Teaching Gains}} \;+\; \underbrace{\Delta \Pi_L(K)}_{\text{Substitution Losses}} \;+\; \underbrace{\Gamma\,\Delta C(K).}_{\text{Non-Cognitive Externality}}
$$

A planner, responding to an accountability framework focused on the first term ($\Delta \Pi_E$), makes choices that are locally rational but socially inefficient because the institutional structure does not compel them to internalize the labor-market effects ($\Delta \Pi_L$) or the harm to non-cognitive skills ($\Gamma \Delta C$). 
This framework clarifies that effective policy must target one or more of these specific wedges to be effective.

Our extensions demonstrate how this core friction may be exacerbated by additional considerations. 
First, the presence of unpriced, non-cognitive skills increases the harm of a naive shift towards skill $A$, creating another source of mismatch rooted in the under-provision of these crucial abilities.
Furthermore, when technology adoption is endogenous, schools systematically over-invest in AI intensity ($\alpha^* > \alpha^\dagger$), guided by an inflated price signal for skill $A$ (Proposition~\ref{prop:ai_cost}). Finally, a non-monotonic, tiered wage structure creates a slightly different problem in the form of a `substitution trap.' 
Here, the risk is not so much over-investment but mis-targeted investment into a low-return skill bracket that an informed planner would strategically avoid, stranding students in an unproductive middle ground (Proposition~\ref{prop:nonmonotonic}).

Our findings could also be framed through the classic lens of comparative advantage. As AI automates routine cognitive tasks (skill type $A$), the comparative advantage of human capital shifts toward areas resistant to automation, such as inter-personal skills ($B$) and non-cognitive abilities ($C$). The policy challenge, therefore, is to realign educational incentives with this evolving frontier of human comparative advantage, rather than simply embracing technology for its immediate productivity benefits in teaching. The non-monotonic extension suggests this frontier may itself be `barbell-shaped,' with human comparative advantage persisting in specific levels of expertise. 

Our findings frame the policy challenge in terms of institutional design: how can we realign educational incentives to track the evolving frontier of human comparative advantage? To address the information wedge, the goal is not perfect foresight but rather to provide educational planners with credible, forward-looking signals about the changing structure of returns, including non-monotonicities like the `substitution trap.' Forward-looking dashboards, 
developed with industry partners, could replace static, high-stakes accountability metrics that are susceptible to Goodhart's Law.

Addressing the incentive wedge created by unpriced non-cognitive skills requires leveraging existing institutional structures. 
Rather than a Pigouvian tax, a more practical approach is to embed incentives for holistic development into accreditation standards and targeted funding. For instance, accrediting bodies could mandate that institutions demonstrate how they cultivate durable skills like resilience, making institutional legitimacy contingent on more than just adopting new technology. This repurposes existing governance mechanisms to reward institutions that develop the human-centric skills least susceptible to automation.

Given the challenges of precise wage forecasting and pricing externalities, direct institutional guardrails offer a reasonable, if imperfect, policy response. 
Such policies could mandate balanced curricula, shifting instructional focus from skills susceptible to AI substitution toward more durable, human-centric abilities. 
They should also govern the adoption of educational AI, evaluating its success based on its holistic impact on student development, including non-cognitive skills like resilience, and not merely on narrow gains in testable outputs. For technical fields with non-monotonic returns, this approach supports a `barbell' strategy: pairing universal digital literacy with selective pathways to deep expertise, thereby helping students avoid the middle-skill `substitution trap.' 
To prevent such regulations from becoming rigid, they must be designed as adaptive rules, for instance, by incorporating review triggers keyed to shifts in skill-specific wage premiums.

The trap we identify is unlikely to be uniform in its effects. 
Low-resource schools may face the strongest institutional pressures to adopt off-the-shelf AI that boosts easily measured skills, potentially exacerbating inequality. 
More subtly, the type of AI deployed may differ systematically: affluent students may be exposed to AI in a human-capital intensive environment that foster metacognition (a complement to skill $C$), while disadvantaged students receive minimal AI tools that provide simple answers, promoting cognitive offloading (a substitute for skill $C$). 
This could create a new digital divide in developmental impact, widening disparities in durable, non-cognitive abilities. 

This paper's theoretical framework generates several testable predictions. 
(1) Using proxies for AI prevalence ($K$), one should find that instructional time in naive institutions (e.g., those with weak industry ties) increasingly diverges from forward-looking indicators of skill demand. 
(2) The adoption of specific AI teaching tools should be highest for skills whose wage premia are forecast to decline. 
(3) Provided that non-cognitive outcomes can be reliably measured, the gap in non-cognitive outcomes between students with high versus low exposure to skill-$A$-focused AI should widen as the real cost ($c$) of the technology falls. 
(4) Optimistically, in settings with tiered returns to $A$, cohort skill distributions should bunch at the thresholds $\underline{A}$ and $\overline{A}$, with a valley in between. 
(5) The over-investment in AI, measured by the gap $\alpha^*(c) - \alpha^\dagger(c)$, widens as the technology's cost ($c$) falls if the naive planner's perceived marginal benefit of adoption is more elastic than the true social marginal benefit (i.e., $|\mathcal{MB}'_{\text{social}}(\alpha^\dagger)| > |\mathcal{MB}'_{\text{naive}}(\alpha^*)|$, per Proposition~\ref{prop:ai_cost}). 
This structural condition is testable via estimation of the underlying production functions $f_A$ and $f_C$.

Our analysis abstracts from general equilibrium effects on capital prices, the rich dynamics of student heterogeneity, and the political economy of school governance, all important areas for future work. While general equilibrium forces can alter magnitudes, the signs of our core comparative statics follow from decreasing-differences conditions and are robust to modest relaxations. 
Our model provides a clear policy principle: preventing students from being trained for obsolescence is an institutional challenge. The governance of educational AI must therefore steer its adoption with forward-looking signals and incentives that reward the holistic development of human capital, guiding students toward their uniquely human comparative advantages.

\appendix

\section{Proofs}

\subsection{Proof of Proposition 1 (Growing Mismatch)}
\label{proof:prop1mismatchK}

\begin{proof}
The result follows directly from the theory of monotone comparative statics.
For the naive planner, the objective function $U_{\text{naive}}(t_A; K)$ has increasing differences in $(t_A, K)$ by assumption (the education channel). Since the constraint set $[0, T]$ is a lattice, Topkis's Theorem implies that the optimal choice set $\tAnaive(K)$ is non-decreasing in $K$. With strict concavity (assumed for uniqueness), $\tAnaive(K)$ is strictly increasing.

For the informed planner, the objective $U_{\text{inf}}(t_A; K)$ exhibits decreasing differences in $(t_A, K)$ because the labor market substitution effect dominates the education channel. Consequently, $\tAinf(K)$ is strictly decreasing in $K$.

The mismatch is defined as $\Mismatch(K) = \tAnaive(K) - \tAinf(K)$. Since the first term is increasing and the second is decreasing, the difference is strictly increasing in $K$.
\end{proof}

\subsection{Proof of Proposition 2 (Non-Cognitive Deficit)}
\label{app:proof_noncog}

\begin{proof}
Let the social objective be $S(t_A, K) = U_{\text{inf}}(t_A, K) + \Gamma C(t_A, K)$.
We have established that $U_{\text{inf}}$ has decreasing differences in $(t_A, K)$.
The non-cognitive production function $C(t_A, K)$ is assumed to be decreasing in $t_A$ (since skill $B$ is more conducive to $C$) and decreasing in $K$ (direct crowding out). Furthermore, the negative impact of AI is exacerbated by allocating time to $A$ (synergistic harm), implying that $C(t_A, K)$ also has decreasing differences.
Since the sum of two functions with decreasing differences also satisfies the property, the social planner's optimal allocation $\tAinfC(K)$ is strictly decreasing in $K$.
Since $\tAnaive(K)$ is increasing, the gap between the naive and social allocations widens.
\end{proof}

\subsection{Proof of Proposition 3 (Over-Adoption)}
\label{proof:prop3_final}

\begin{proof}
The naive planner equates marginal benefit to marginal cost: $\mathcal{MB}_{\text{naive}}(\alpha) = c$.
The social planner does the same: $\mathcal{MB}_{\text{social}}(\alpha) = c$.
The naive planner's marginal benefit is higher for two reasons:
1. $p_A > w_A(K)$ (optimism about skill value).
2. They ignore the negative term $\Gamma \frac{\partial C}{\partial \alpha}$ (neglect of non-cognitive harm).
Thus, for any $\alpha$, $\mathcal{MB}_{\text{naive}}(\alpha) > \mathcal{MB}_{\text{social}}(\alpha)$.
Assuming diminishing marginal returns, this implies the naive planner chooses a strictly higher intensity $\alpha^* > \alpha^\dagger$.
As $c$ falls, both agents increase adoption, but the difference in slopes (driven by the additional curvature of the non-cognitive penalty) ensures the gap widens.
\end{proof}

\subsection{Proof of Proposition 4 (The Substitution Trap)}
\label{app:proof_prop4}

This proposition relies on the non-monotonic wage structure introduced in Section \ref{sec:extension-nonmonotonic}. Here we provide the formal details.

\noindent \emph{Wage Structure and Assumptions.}\\
We model the return to skill $A$ as a piecewise lump-sum payment, $w_A(A;K)$. Let there be two skill thresholds, $0 < \underline{A} < \overline{A}$, such that:
\begin{equation}
\label{eq:piecewise-wA}
w_A\bigl(A; K\bigr)
\;=\;
\begin{cases}
    \underline{w}_A(K), & 0 \;\le\; A < \underline{A} \quad \text{(basic skill `literacy')}\\[4pt]
    \widetilde{w}_A(K), & \underline{A} \;\le\; A < \overline{A} \quad \text{(AI-substitution trap)}\\[4pt]
    \overline{w}_A(K),  & A \;\ge\; \overline{A} \quad \text{(advanced expertise)}
\end{cases}
\end{equation}
We assume $\overline{w}_A(K) > \underline{w}_A(K) \ge \widetilde{w}_A(K)$, reflecting the ``substitution trap'' in the middle range. We further assume that for any $K$, the thresholds are reachable at unique time allocations $\tau_{\underline{A}}(K)$ and $\tau_{\overline{A}}(K)$, which are weakly decreasing in $K$ as AI enhances learning efficiency.

The proof proceeds by characterizing the optimal choices of the naive and informed planners separately and then comparing them.

\begin{proof}
\textbf{(i) The Naive Planner.}
The naive planner maximizes $U_{\text{naive}}(t_A; K) = p_A f_A(t_A, \gK) + p_B f_B(T-t_A, \gK)$. Since $f_A$ and $f_B$ are strictly concave and the objective satisfies increasing differences in $(t_A, K)$ (as assumed in the education channel), the optimal allocation $\tAnaive(K)$ is unique and strictly increasing in $K$. The naive planner, reacting to the linear price signal $p_A$, smoothly increases time investment as AI makes skill $A$ easier to acquire.

\textbf{(ii) The Informed Planner.}
The informed planner maximizes $U_{\inf}(t_A; K) = w_A(f_A(t_A, \gK); K) + w_B(K) f_B(T-t_A, \gK)$.
Because $w_A$ is constant within each skill tier while the opportunity cost (lost skill $B$) strictly increases with $t_A$, the objective function is strictly decreasing within the interior of any tier. Thus, the optimal allocation must be at the lower bound of a tier: $\tAinf(K) \in \{0, \tau_{\underline{A}}(K), \tau_{\overline{A}}(K)\}$.

Comparing these candidates, the ``substitution trap'' option $\tau_{\underline{A}}(K)$ yields a lower wage ($\widetilde{w}_A \le \underline{w}_A$) and lower skill $B$ production than the option $t_A=0$ (basic literacy). Therefore, $\tau_{\underline{A}}(K)$ is strictly dominated. The informed planner's choice reduces to a ``barbell'' strategy: either $t_A=0$ (basic literacy) or $t_A=\tau_{\overline{A}}(K)$ (advanced expertise).

\textbf{(iii) Mismatch Dynamics.}
The mismatch is defined as $\Mismatch(K) = \tAnaive(K) - \tAinf(K)$.
\begin{itemize}
    \item If the informed planner maintains the strategy $\tAinf(K)=0$, the mismatch $\tAnaive(K) - 0$ grows monotonically because $\tAnaive(K)$ is strictly increasing.
    \item If the informed planner targets advanced expertise ($\tAinf(K)=\tau_{\overline{A}}(K)$), the mismatch is $\tAnaive(K) - \tau_{\overline{A}}(K)$. Since $\tAnaive(K)$ is increasing and the cost of expertise $\tau_{\overline{A}}(K)$ is decreasing (due to AI efficiency), the mismatch again grows.
    \item However, if the wage premium for expertise fluctuates such that the informed planner switches strategies (e.g., from $\tau_{\overline{A}}(K)$ to $0$), $\tAinf(K)$ jumps discontinuously. This causes the mismatch function to exhibit non-monotonic jumps.
\end{itemize}
Thus, while the naive planner smoothly steers students into the intermediate ``trap,'' the informed planner avoids it, leading to a potentially non-monotonic divergence in allocations.
\end{proof}

\section{Pilot survey evidence on AI-teachable skills and workplace disruption}
\label{app:pilot}

A preregistered pilot survey explores the relationship between (i) O*NET skills that educators view as relatively easy to teach using AI tools and (ii) an index of potential AI-driven disruption at the skill level. Given the small and non-representative sample and the use of LLM-derived disruption measures, the results should be interpreted as suggestive rather than as core quantitative evidence.

For further details on the empirical motivation, including the full set of survey instruments, pre-registration documents, and additional robustness checks, please refer to our OSF repository: \url{https://osf.io/nwy4c/?view_only=25b2a80ac79a44b890284dde2d70b8b8}.

\begin{figure}[!h] 
  \begin{center}
        \includegraphics[width=0.6\textwidth]{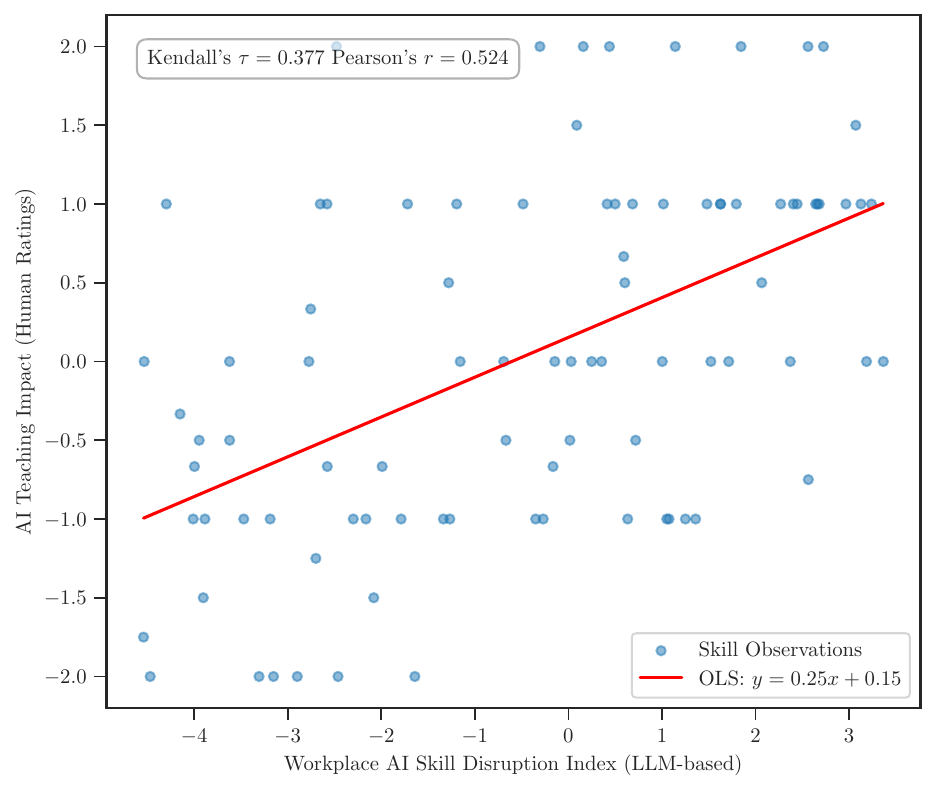}  
  \end{center}
  \caption{Correlation between Perceived AI Teaching Impact and AI Skill Disruption}  \label{fig:scatter_ai_teach_vs_sub_appendix}
   \vspace{10pt}
\begin{minipage}{0.95\textwidth}
\subcaption*{\textit{Notes:} Each point represents one of 90 unique skills.
The y-axis shows the mean rating from our survey (`AI Teaching Impact'); the x-axis shows our LLM-derived workplace `AI Skill Disruption Index.' The line shows the OLS fit (however inference is based on Kendall's $\tau_b$, not the OLS slope.) The pre-specified Kendall's $\tau_b$ rank correlation for these data is $0.377$, $p<0.001$.}
\end{minipage}
\end{figure}

\clearpage

\bibliographystyle{econometrica}
\bibliography{ai_ed}

\end{document}